\documentclass[onecolumn,11pt]{article}
\usepackage[top=1in, bottom=1in, left=1in, right=1in]{geometry}
\usepackage{amsmath,amsfonts,amscd,amssymb}
\usepackage{graphicx}
\usepackage{epstopdf}
\usepackage{overpic}
\usepackage{cancel}
\usepackage{rotating}
\usepackage{url}
\usepackage{caption}
\usepackage{color}
\usepackage{rotating}
\usepackage{multirow}
\usepackage{wrapfig}
\usepackage{mathtools}
\usepackage{subeqnarray}
\usepackage{setspace}
\usepackage{palatino} 
\usepackage[authoryear,round,sort&compress]{natbib}
\renewcommand{\citet}[1]{\cite{#1}}%

\renewcommand{\b}[1]{\boldsymbol{#1}}
\newcommand{\B}[1]{\mathbf{#1}}
\newcommand{\ii}[0]{\mathrm{i}}
\newcommand{\Rey}[0]{Re}

\usepackage[bottom,flushmargin,hang,multiple]{footmisc}
\usepackage{lipsum}
\newcommand\blfootnote[1]{%
  \begingroup
  \renewcommand\thefootnote{}\footnote{#1}%
  \addtocounter{footnote}{-1}%
  \endgroup
}

\definecolor{header1}{cmyk}{0,0,0,1}

\title{\huge{\vspace{-.5in}\textbf{Data-driven resolvent analysis}}\vspace{.175in}\\

\normalsize{Benjamin Herrmann$^{12*}$, Peter J. Baddoo$^3$, Richard Semaan$^2$,\\ Steven L. Brunton$^1$, and Beverley J. McKeon$^4$}\vspace{.1in}\\
\footnotesize{$^1$ Department of Mechanical Engineering, University of Washington, Seattle, WA 98195, USA}\\
\footnotesize{$^2$ Institute of Fluid Mechanics, Technische Universit\"at Braunschweig, 38108 Braunschweig, Germany}\\
\footnotesize{$^3$ Department of Mathematics, Imperial College London, South Kensington Campus, London SW7 2AZ, UK}\\
\footnotesize{$^4$ Graduate Aerospace Laboratories, California Institute of Technology, Pasadena CA 91125,
USA}\vspace{-.175in}
}
\date{}
\begin{document}
\maketitle

\blfootnote{$^*$ Corresponding author (benherrm@uw.edu).}

\vspace{-.4in}
\begin{abstract}
Resolvent analysis identifies the most responsive forcings and most receptive states of a dynamical system, in an input--output sense, based on its governing equations.
Interest in the method has continued to grow during the past decade due to its potential to reveal structures in turbulent flows, to guide sensor/actuator placement, and for flow control applications. 
However, resolvent analysis requires access to high-fidelity numerical solvers to produce the linearized dynamics operator. 
In this work, we develop a purely data-driven algorithm to perform resolvent analysis to obtain the leading forcing and response modes, without recourse to the governing equations, but instead based on snapshots of the transient evolution of linearly stable flows.
 The formulation of our method follows from two established facts: $1)$ dynamic mode decomposition can approximate eigenvalues and eigenvectors of the underlying operator governing the evolution of a system from measurement data, and $2)$ a projection of the resolvent operator onto an invariant subspace can be built from this learned eigendecomposition.
 We demonstrate the method on numerical data of the linearized complex Ginzburg--Landau equation and of three-dimensional transitional channel flow, and discuss data requirements.
 The ability to perform resolvent analysis in a completely equation-free and adjoint-free manner will play a significant role in lowering the barrier of entry to resolvent research and applications.
\end{abstract}

\section{Introduction}
\label{sec:intro}

The resolvent is a linear operator that governs how harmonic forcing inputs are amplified by the linear dynamics of a system and mapped onto harmonic response outputs. 
Resolvent analysis refers to the inspection of this operator to find the most responsive inputs, their gains, and the most receptive outputs. 
The resulting low-rank approximation of the forcing-response dynamics of the full system is extremely valuable for modeling, controlling, and understanding the physics of fluid flows. 
Interest in the approach has continued to grow since \citet{McKeon2010b} showed that, by interpreting the nonlinear term in the Fourier-transformed Navier--Stokes equations as an exogenous harmonic forcing, resolvent analysis can uncover elements of the structure in wall turbulence.

Two decades before coined as such, resolvent analysis was first used in the seminal work of \citet{Trefethen1993} to study the response of linearly stable flows to deterministic external disturbances, such as those coming from wall roughness, acoustic perturbations, body forces, or free-stream turbulence. 
A key result was to identify the non-normality of the linearized operator as the cause of transient energy amplification of disturbances, even for cases deemed as stable by eigenvalue analysis. 
During the $1990$s, nonmodal stability theory emerged to provide a more complete picture of the linear perturbation dynamics for fluid flows using an initial-value problem formulation, as a complement to the eigenproblem from classic hydrodynamic stability theory~\citep{Schmidbook,Schmid2007,Schmid2014a}. 
This formulation allowed the study of the response of fluid flows to initial conditions~\citep{Gustavsson1991,Butler1992a,Farrell1993a,Reddy1993a,Hwang2010,Herrmann2018b}, stochastic inputs~\citep{Farrell1993b,DelAlamo2006,Hwang2010a,Hwang2010}, and harmonic forcing~\citep{Jovanovic2005,Hwang2010,Hwang2010a,Herrmann2018b}. 
Another landmark is the framework adopted by~\citet{Jovanovic2005} that focuses on the response of certain outputs of interest to forcing of specific input components. 
This input--output viewpoint allows the examination of localized disturbances and provides mechanistic insight into multi-physics systems~\citep{Jeun2016,Herrmann2018,Jovanovic2020}, making it particularly relevant for control applications. 
Reduced-order models based on resolvent modes have been studied for turbulent channel flows~\citep{Moarref2013,Moarref2014,McKeon2017a,Abreu2020}, laminar and turbulent cavity flows~\citep{Gomez2016,Sun2020}, and turbulent jets~\citep{Schmidt2018a,Lesshafft2019}. 
Other exciting recent developments include the design of an airfoil separation control strategy based on resolvent analysis by~\citet{Yeh2019}, the harmonic resolvent formalism to capture cross-frequency interactions in periodic flows by~\citet{Padovan2020}, and the application of a nonlinear input--output analysis to boundary layer transition by~\citet{Rigas2020}. 

Even though there is a growing interest in the community to study flows with two and three inhomogeneous directions, global resolvent analysis is still far from commonplace. 
The main reasons for this are the requirement of a high-fidelity solver for the linearized governing equations, and the computational cost and memory allocation associated with handling a very large operator. 
The latter challenge has been addressed using matrix-free iterative techniques~\citep{Bagheri2009a,Monokrousos2010,Loiseau2019,Martini2020}. 
However, this approach requires having access to a high fidelity solver for the adjoint equations, which adds to the first challenge. 
A promising alternative, first used by~\citet{Moarref2013} and further investigated by~\citet{Ribeiro2020}, is the use of randomized numerical linear algebra techniques~\citep{Liberty2007,Halko2011}. 
To simultaneously addresss both issues, we propose a purely data-driven approach to obtain the resolvent operator that doesn't rely on access to the governing equations and can be combined with randomized methods to alleviate the computational expense if needed.

The unprecedented availability of high-fidelity numerical simulations and experimental measurements has lead to incredible growth of research in data-driven modeling of dynamical systems during the past decade~\citep{Brunton2016,Rudy2017,Loiseau2018,Raissi2019jcp,Blanchard2019,databook,Li2020,raissi2020science}. 
In fluid dynamics, this has led to the development and application of machine learning (ML) algorithms to extract dominant coherent structures from flow data~\citep{Taira2017aiaa,Taira2019,Brenner2019prf,Brunton2020}. 
Dynamic mode decomposition (DMD) is a particularly relevant technique introduced by~\citet{Schmid2010} to learn spatiotemporal patterns from time-resolved data that are each associated with a single frequency and growth/decay rate~\citep{Kutz2016book}. 
During the last decade, considerable effort has been devoted to interpret its application to nonlinear dynamical systems, based on a deep connection to Koopman theory~\citep{Rowley2009,Mezic2013}, and to develop numerous extensions to allow the use of non-sequential measurements~\citep{Tu2014}, promote sparsity in the solution~\citep{Jovanovic2014}, work with streaming datasets~\citep{Hemati2014}, improve its accuracy with nonlinear observables~\citep{Williams2015}, incorporate the effect of control inputs~\citep{Proctor2016}, add robustness using nonlinear optimization~\citep{Askham2018}, and enable its application to massive datasets with randomized linear algebra techniques~\citep{Erichson2019}. 
Recently, other data-driven modal decompositions have been derived, such as spectral proper orthogonal decomposition (SPOD)~\citep{Lumley1970,Towne2018}, and the bispectral mode decomposition (BMD)~\citep{Schmidt2020}. 
\citet{Towne2018} showed that, for turbulent and statistically stationary flows, SPOD modes are equivalent to the output resolvent modes if the nonlinear forcing exhibits no preferential direction. 
Nonetheless, it is the input resolvent modes that provide insight into the most amplified flow structures, the most sensitive actuator locations, and the most responsive control inputs.

\begin{figure}
\centering
\includegraphics[width=1\textwidth]{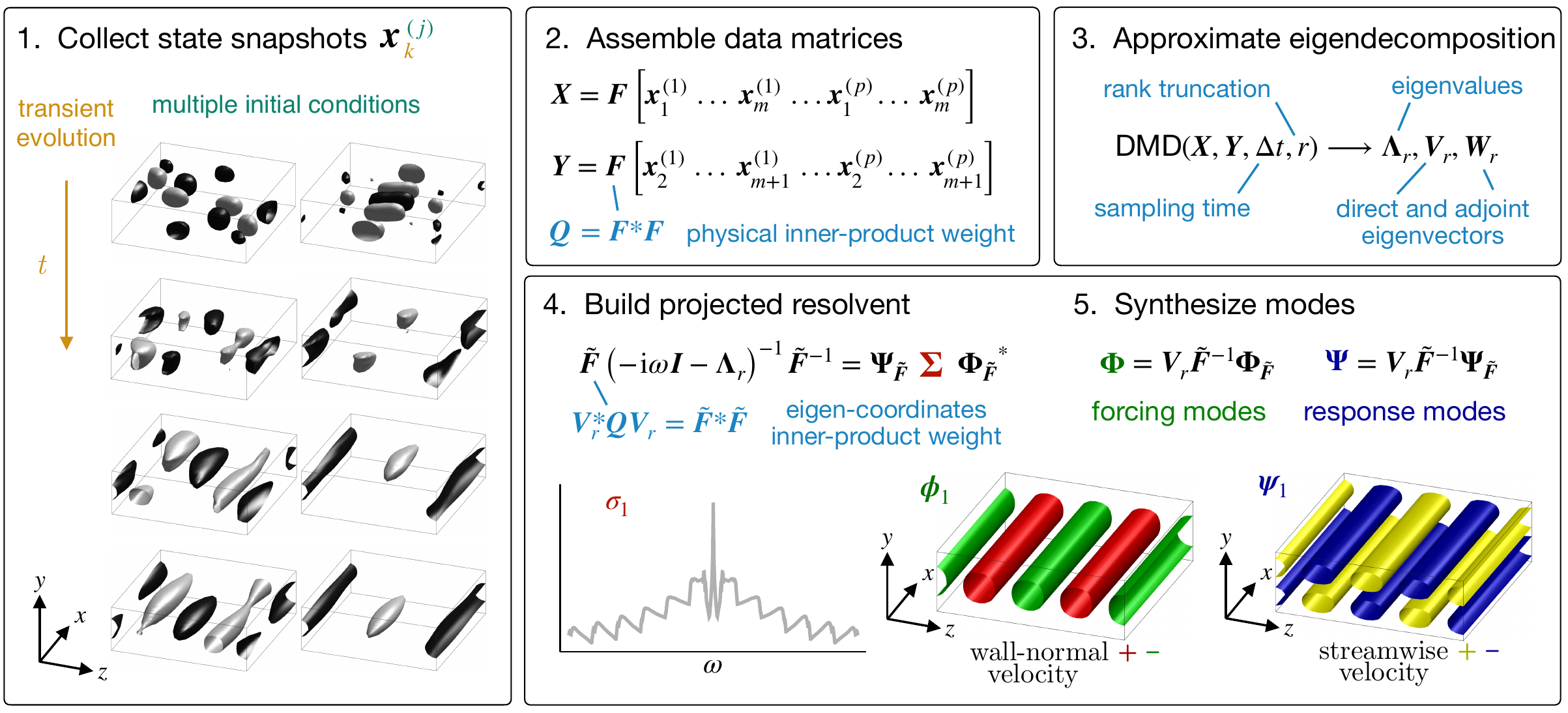}
\caption{Schematic of the data-driven resolvent analysis algorithm demonstrated on the transitional channel flow example detailed in \S\ref{sec:results2}. Data is collected from time recordings of the system of interest, where one or more initial conditions are used to generate transient dynamics. Measurements are stacked into data matrices that are used to approximate an eigendecomosition of the underlying system via dynamic mode decomposition. A projection of the resolvent operator onto the span of the learned eigenvectors is analyzed, and, finally, the produced modes are lifted to physical coordinates.}
\label{method}
\end{figure}

In this work, we present an algorithm to obtain the leading input and output resolvent modes, and the associated gains, of linearly stable flows directly from data by following the procedure shown in Figure~\ref{method}. 
The method relies on DMD to approximate the eigenvalues and eigenfunctions of the system based on snapshots from one or more transient trajectories of the flow. 
We show that, using an appropriate inner-product, the resolvent of the matrix of DMD eigenvalues is the resolvent of the system projected onto the span of the DMD eigenvectors, which are subsequently used to synthesize the resolvent modes in physical coordinates. 
Our method is able to find the optimal forcing, response, and gain of a dynamical system in an equation-free manner, and without access to data from adjoint simulations, therefore opening up the possibility of resolvent analysis purely based on experimental measurements.

The remainder of the paper is organized as follows. 
A brief and practical formulation of resolvent analysis is presented in \S\ref{sec:background}, followed by a general description of our proposed method in \S\ref{sec:method}. 
We demonstrate the approach on two examples and discuss data requirements and convergence in \S\ref{sec:results}. 
Our conclusions and thoughts on possible extensions are offered in \S\ref{sec:conclusions}.

\section{Resolvent analysis}
\label{sec:background}

In this section, we present a brief and practical formulation of resolvent analysis, followed by a description of a dimensionality reduction approach first used in the context of nonmodal stability theory by~\citet{Reddy1993a}.

Let us consider a forced linear dynamical system
\begin{equation}
\dot{\b{x}}=\B{A}\b{x}+\b{f}, \label{sys}
\end{equation}
where the dot denotes time-differentiation, $\b{x}\in \mathbb{C}^n$ is the state vector, $\B{A} \in \mathbb{C}^{n\times n}$ is the linear dynamics matrix, and $\b{f}\in \mathbb{C}^n$ is an external driving force. 
Such a system may arise from a semi-discretized partial differential equation, and in the case of fluid flows, the incompressible Navier--Stokes equations can be written in this form by projecting the velocity field onto a divergence-free basis to eliminate the pressure variable. 
The state $\b{x}$ may either represent the deviation from a steady state of a laminar flow, or fluctuations about the temporal mean of a statistically stationary unsteady flow. 
In both cases the matrix $\B{A}$ is the linearization of the underlying nonlinear system about the corresponding base flow, either the equilibrium or the mean. 
However, if we are dealing with an unsteady flow, it is important to note that a Reynolds decomposition yields a nonlinear term that typically cannot be neglected. 
In this scenario, the nonlinearity is lumped into $\b{f}$ and considered as an external forcing, with the caveat that the system may exhibit preferred input directions due to internal feedback between linear amplification and nonlinear interactions~\citep{McKeon2010b,Beneddine2016}.

In this work, we will focus on the case where $\b{x}$ is the deviation from a stable steady state and $\b{f}$ is an exogenous input with no preferential direction, representing disturbances from the environment, model discrepancy, or an open-loop control actuation. For a harmonic forcing ${\b{f}(t) = \hat{\b{f}}e^{-\ii\omega t} +\text{c.c.}}$, where $\text{c.c.}$ means complex conjugate, $\omega\in \mathbb{R}$ is the angular driving frequency and $t\in \mathbb{R}$ the time variable, the long-term response is also harmonic, $\b{x}(t) = \hat{\b{x}}e^{-\ii\omega t} +\text{c.c.}$, and is governed by the particular solution to \eqref{sys} given by
\begin{equation}
\hat{\b{x}} = \left(-\ii\omega\B{I}-\B{A}\right)^{-1}\hat{\b{f}},
\end{equation}
where $\B{I}$ is the $n\times n$ identity matrix. Let $\B{H}(\omega)=\left(-\ii\omega\B{I}-\B{A}\right)^{-1}$ be the matrix approximation of the resolvent operator, where the negative sign accompanying the $\omega$-term follows the convention used for travelling waves. We seek the largest input--output gain, $\sigma_1(\omega)$, optimized over all possible forcing vectors $\hat{\b{f}}$, or more formally
\begin{equation}
\sigma_1(\omega)=
\max_{\hat{\b{f}}\ne\b{0}}\frac{\| \hat{\b{x}}\|^2_{\B{Q}}}{\| \hat{\b{f}}\|^2_{\B{Q}}}=
\max_{\hat{\b{f}}\ne\b{0}}\frac{\| \B{H}(\omega)\hat{\b{f}}\|^2_{\B{Q}}}{\| \hat{\b{f}}\|^2_{\B{Q}}},\label{s1}
\end{equation}
where $\| \hat{\b{x}} \|^2_\B{Q}=\hat{\b{x}}^*\B{Q}\hat{\b{x}}$, with $( \ )^*$ denoting the Hermitian transpose, measures the size of the state based on a physically meaningful metric given by the positive-definite weighting matrix $\B{Q}$. This weighting accounts for integration quadratures, non-uniform spatial discretizations, and appropriate scaling of heterogeneous variables in multi-physics systems~\citep{Jeun2016,Herrmann2018}. The Cholesky decomposition is used to factorize $\B{Q}=\B{F}^*\B{F}$ and relate the physically-relevant norm to the standard Euclidean $2$-norm via $\| \hat{\b{x}} \|^2_{\B{Q}}=\| \B{F}\hat{\b{x}} \|^2_2\,$. With this scaling of the states, and using the definition of the vector-induced matrix norm, the optimal gain optimization problem \eqref{s1} is equivalent to
\begin{equation}
\sigma_1(\omega)=\| \B{F}\B{H}(\omega)\B{F}^{-1}\|^2_2.\label{s2}
\end{equation}
The solution to \eqref{s2}, along with a hierarchy of optimal and suboptimal forcing and response vectors, is given by the singular value decomposition (SVD) of the weighted resolvent
\begin{equation}
\B{F}\B{H}(\omega)\B{F}^{-1}=\b{\Psi}_{\B{F}}(\omega)\b{\Sigma}(\omega)\b{\Phi}_{\B{F}}^*(\omega),\label{svd}
\end{equation}
where $\b{\Sigma} \in \mathbb{R}^n$ is a diagonal matrix containing the gains $\sigma_1 \geq \sigma_2 \geq \dotsc\geq \sigma_n \geq 0$, also known as singular values, and $\b{\Phi}_{\B{F}}=\B{F} \ [\b{\phi}_1 \ \b{\phi}_2\ \dotsc \ \b{\phi}_n]\in \mathbb{C}^{n\times n}$ and $\b{\Psi}_{\B{F}}=\B{F}[\b{\psi}_1 \ \b{\psi}_2\ \dotsc \ \b{\psi}_n]\in \mathbb{C}^{n\times n}$ are unitary matrices whose columns, when left-multiplied by $\B{F}^{-1}$, yield the input and output resolvent modes, $\b{\phi}_j$ and $\b{\psi}_j$, respectively.

\section{Data-driven resolvent analysis}
\label{sec:method}

Our data-driven approach to perform resolvent analysis relies on dynamic mode decomposition (DMD) to approximate the eigenvalues and eigenvectors of the underlying dynamical system. 
Among the many choices of DMD variants, we use the \emph{exact} DMD approach derived by~\citet{Tu2014}. 
In the absence of forcing, the evolution of measurements of the dynamical system of interest \eqref{sys} is governed by
\begin{equation}
\b{x}_{k+1}=\exp(\B{A}\Delta t)\b{x}_k , \label{sysd}
\end{equation}
where $\b{x}_k$ is the measurement at time $t_k=k\Delta t$, and $\Delta t$ is the sampling time. As in the previous section, let the the weight matrix $\B{Q}=\B{F}^*\B{F}$ define a physically meaningful norm to quantify the size of the state vector. The following transformation allows us to work in the $2$-norm framework,
\begin{equation}
\B{F}\b{x}_{k+1}=\B{F}\exp(\B{A}\Delta t)\B{F}^{-1}\B{F}\b{x}_k=\b{\Theta}\B{F}\b{x}_k, \label{sysdF}
\end{equation}
where $\b{\Theta}=\B{F}\exp(\B{A}\Delta t)\B{F}^{-1}$ evolves the weighted measurements $\B{F}\b{x}_k$ one time step into the future. Under this transformation, the adjoint of $\b{\Theta}$ based on the $\B{Q}$-norm is equivalent to the Hermitian adjoint. 
Thus, using the weighted measurements, we can proceed using readily available DMD codes. 
To begin, we collect snapshots of the state denoted by $\b{x}_k^{(j)}$, where the subscript $k\in\lbrace 1, \ \dotsc, \ m+1\rbrace$ denotes the sample number, and the superscript $j\in\lbrace 1, \ \dotsc, \ p\rbrace$ denotes different trajectories started from $p \geq 1$ initial conditions. 
The next step is to assemble the weighted data matrices
\begin{subequations}\label{data}
\begin{align}
\B{X} &= \B{F}\left[\b{x}^{(1)}_1 \ \b{x}^{(1)}_2... \ \b{x}^{(1)}_m \ \left| \ \b{x}^{(2)}_1 \ \b{x}^{(2)}_2... \ \b{x}^{(2)}_m \ \right| \dotsm \ \left|  \ \b{x}^{(p)}_1 \ \b{x}^{(p)}_2... \ \b{x}^{(p)}_m\right], \right.\\
\B{Y} &= \B{F}\left[\b{x}^{(1)}_2 \ \b{x}^{(1)}_3... \ \b{x}^{(1)}_{m+1} \ \left| \ \b{x}^{(2)}_2 \ \b{x}^{(2)}_3... \ \b{x}^{(2)}_{m+1} \ \right| \dotsm \ \left|  \ \b{x}^{(p)}_2 \ \b{x}^{(p)}_3... \ \b{x}^{(p)}_{m+1}\right], \right.
\end{align}
\end{subequations}
where $\B{X}$ and $\B{Y} $ are of size $n\times p m$. Based on these data matrices, the DMD framework with a rank-$r$ truncated SVD yields the matrices $\B{D}_r= \text{diag}(\rho_1,\rho_2,\dotsc,\rho_r)\in \mathbb{C}^{r\times r}$ containing the approximated eigenvalues, and $\B{V}_{r,\B{F}}\in \mathbb{C}^{n\times r}$ and $\B{W}_{r,\B{F}}\in \mathbb{C}^{n\times r}$, whose columns are the approximated direct and adjoint eigenvectors of the underlying operator $\b{\Theta}$. 
The reader is referred to the work of~\citet{Tu2014} for the derivation of the adjoint DMD modes. 
These eigenvalues and vectors are related to those corresponding to $\B{A}$ via $\lambda_j=\log(\rho_j)/\Delta t$, $\B{V}_r=\B{F}^{-1}\B{V}_{r,\B{F}}$, and $\B{W}_r=\B{F}^{-1}\B{W}_{r,\B{F}}$. 
Hence, we obtain $\b{\Lambda}_r= \text{diag}(\lambda_1,\lambda_2,\dotsc,\lambda_r)\in \mathbb{C}^{r\times r}$, $\B{V}_r= \ [\b{v}_1 \ \b{v}_2\dotsc \ \b{v}_r]\in \mathbb{C}^{n\times r}$, and ${\B{W}_r= \ [\b{w}_1 \ \b{w}_2\dotsc \ \b{w}_r]\in \mathbb{C}^{n\times r}}$, that satisfy
\begin{subequations}\label{eig}
\begin{align}
\B{A}\B{V}&=\B{V}\b{\Lambda},\\
\B{A}^+\B{W}&=\B{W}\b{\Lambda}^*,
\end{align}
\end{subequations}
for an unknown underlying operator $\B{A}$, where $\B{A}^+=\B{Q}^{-1}\B{A}^*\B{Q}$ is its $\B{Q}$-norm adjoint. In other words, we use DMD as a data-driven eigendecomposition, which is not surprising considering that its connection to Arnoldi methods and Krylov subspaces has been clear since the origins of the algorithm \citep{Schmid2010}.

Next, we seek an approximation of the resolvent operator built on $\b{\Lambda}_r$, $\B{V}_r$, and $\B{W}_r$. 
Our approach leverages an operator-based dimensionality reduction technique first used in the context of nonmodal stability analysis by \citet{Reddy1993a}. 
We now return our attention to the forced system \eqref{sys}, and consider an eigenvector expansion of $\b{x}$ and $\b{f}$, as follows
\begin{subequations}\label{sysred}
\begin{align}
\b{x}(t)&=\B{V}_r\b{a}(t),\\
\b{f}(t)&=\B{V}_r\b{b}(t),
\end{align}
\end{subequations}
where $\b{a}= \ [a_1 \ a_2\ \dotsc \ a_r]^T\in \mathbb{C}^r$, and $\b{b}= \ [b_1 \ b_2\ \dotsc \ b_r]^T\in \mathbb{C}^r$ are the vector of expansion coefficients in eigenvector coordinates. In the work of \citet{Reddy1993a} $\B{V}_r$ are the eigenvectors associated with the first $r$ eigenvalues with largest real part of a known operator $\B{A}$, whereas here they are the DMD modes. Substitution of \eqref{sysred} in \eqref{sys}, and taking the inner product with $\B{W}_r$ at both sides yields
\begin{equation}
\dot{\b{a}}=\b{\Lambda}_r\b{a} + \b{b},
\end{equation} 
where we have used the bi-orthogonality property between the sets of direct and adjoint eigenvectors, and assumed that they have been normalized such that $\B{W}_r^*\B{Q}\B{V}_r=\B{I}\in \mathbb{R}^{r\times r}$. 
Because we are now working in different coordinates, if we want to retain the physical meaning of the norm, we need to adjust our inner-product accordingly. 
The new weighting matrix is derived as $\|\b{x}\|_{\B{Q}}^2 = \b{x}^*\B{Q}\b{x}=\b{a}^*\B{V}_r^*\B{Q}\B{V}_r\b{a}=\|\tilde{\B{F}}\b{a}\|_2^2$, where we have defined the new matrix $\tilde{\B{F}}\in \mathbb{C}^{r \times r}$ from the Cholesky factorization of $\B{V}_r^*\B{Q}\B{V}_r=\tilde{\B{F}}^*\tilde{\B{F}}$. 
We are now ready to proceed with the resolvent analysis for the system \eqref{sysred}. 
As presented in the previous section, the weighted resolvent modes and gains are obtained from the SVD of
\begin{equation}
\tilde{\B{F}}(-\ii\omega\B{I}-\b{\Lambda}_r)^{-1}\tilde{\B{F}}^{-1}=\b{\Psi}_{\tilde{\B{F}}}\b{\Sigma}\b{\Phi}_{\tilde{\B{F}}}^*.\label{svdred}
\end{equation}
The final step is to synthesize the resolvent modes in physical coordinates, as follows
\begin{subequations}\label{modes}
\begin{align}
\b{\Phi}&=\B{V}_r\tilde{\B{F}}^{-1}\b{\Phi}_{\tilde{\B{F}}},\\
\b{\Psi}&=\B{V}_r\tilde{\B{F}}^{-1}\b{\Psi}_{\tilde{\B{F}}}.
\end{align}
\end{subequations}
A schematic that summarizes the entire procedure is shown in Figure \ref{method}. 
It is worth pointing out that an analogous procedure can be carried out to obtain data-driven transient growth modes, simply by replacing the resolvent operator with the matrix exponential propagator evaluated at a finite time-horizon. 
In addition, notice that the reduced-order resolvent matrix in \eqref{svdred} is of size $r\times r$, meaning that its full SVD requires $\mathcal{O}(r^3)$ operations instead of $\mathcal{O}(n^3)$, and therefore is considerably cheaper to compute. 
This projection onto the span of the eigenvectors has been successfully used in operator-based frameworks to achieve computational speedups of a few orders of magnitude for non-modal stability analysis \citep{Reddy1993a,Herrmann2018}, and several orders of magnitude for pseudospectra computations \citep{Trefethen2005}. 
Although the potential time savings are promising, the computational bottleneck of the presented data-driven method is the truncated SVD in the DMD step, which has an operation count of $\mathcal{O}(nl^2m^2)$. 
However, our approach benefits from all past and future innovations to improve the accuracy, robustness, flexibility, and speed of DMD~\citep{Kutz2016book}.

\section{Examples and discussion}
\label{sec:results}

In this section, we demonstrate the application of data-driven resolvent analysis on two example problems.

\subsection{Complex Ginzburg--Landau equation}
\label{sec:results1}

Our first example is the linearized complex Ginzburg--Landau equation, which is a typical model for instabilities in spatially-evolving flows. The system is governed by the linear operator
\begin{equation}
\B{A} = -\nu \B{D}_x + \gamma \B{D}_x^2 + \mu(x),
\end{equation}
where $x$ is the spatial coordinate, and $\B{D}_x$ and $\B{D}^2_x$ are the $1^{st}$ and $2^{nd}$-order spatial differentiation matrices with homogeneous boundary conditions at $x\rightarrow \pm \infty$. 
We choose a quadratic spatial dependence for the parameter $\mu(x)=(\mu_0-c_{\mu}^2)+\tfrac{\mu_2}{2}x^2$, that has been used previously by several authors \citep{Bagheri2009b, Chen2011,Towne2018}. 
The other parameters are set to $\mu_0=0.23$, $\mu_2=-0.01$, $\nu=2+0.4\ii$, and $\gamma=1-\ii$, giving rise to linearly stable dynamics. 
As in \citet{Bagheri2009b}, we use spectral collocation based on Gauss-weighted Hermite polynomials to build the differentiation matrices $\B{D}_x$ and $\B{D}^2_x$ and the integration quadrature $\B{Q}$ \citep{Weideman2000}. 
The spatial coordinate is discretized into $n=220$ collocation points, and the domain is truncated to $x\in [-85,85]$, which is sufficient to enforce the far-field boundary conditions.

\begin{figure}
\centering
\includegraphics[width=1\textwidth]{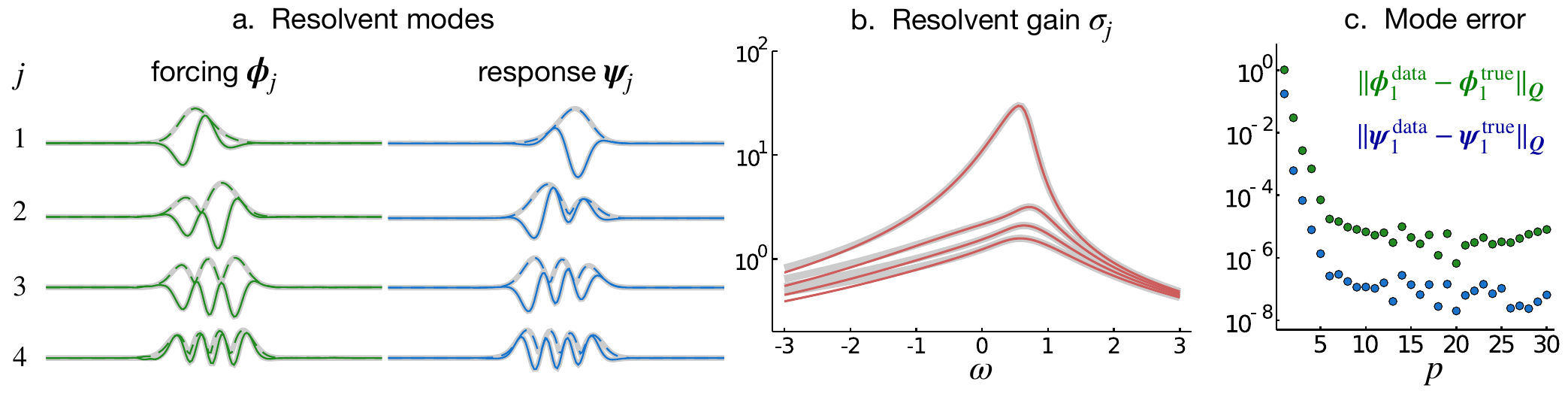}
\caption{Data-driven resolvent analysis of the linearized complex Ginzburg--Landau equation. \textit{a}.~The first four forcing and response modes, where solid and dashed lines show the real part and magnitude of the modes. \textit{b}.~Their respective gains as a function of frequency. In \textit{a} and \textit{b}, the thick gray lines show operator-based quantities for a ground-truth comparison. \textit{c}. $\B{Q}$-norm error between the operator-based and the data-driven leading resolvent modes as a function of the number of trajectories $p$ considered in the dataset.}
\label{gl}
\end{figure}

Data is generated from $30$ simulations that are each started from different initial conditions which we choose to be the first $30$ Gauss-weighted Hermite polynomials. 
We record $m=100$ snapshots that are sampled every $\Delta t=0.5$ time units. 
Data-driven resolvent analysis is performed using snapshot matrices assembled considering measurements from the first trajectory only. 
Subsequently, the method is applied on snapshot matrices where measurements from the other simulations are sequentially concatenated one by one, to investigate the convergence behavior in regards to the amount of data required. The $\B{Q}$-norm error between the operator-based and the data-driven leading resolvent modes as a function of the number of trajectories considered is shown in Figure~\ref{gl}. 
As more data is included, the abrupt drop in the error is expected, since DMD is able to accurately approximate a larger number of eigenvectors, therefore enriching the basis we use to represent the resolvent modes. 
The first four resolvent modes, as well as their gains as a function the input frequency are shown in Figure~\ref{gl}. 
For all results presented in this section, we used $r=24$ for the rank truncation in both DMD and the data-driven eigenbasis $\B{V}_r$. 
It is worth pointing out that, caution is needed when retaining more vectors in $\B{V}_r$ to not include spurious eigenvectors, which instead of enriching the basis can be detrimental to the performance of the method.

\subsection{Transitional channel flow}
\label{sec:results2}

Our second example is the three-dimensional flow in a plane channel of finite length and depth, and with periodic streamwise and spanwise boundary conditions. 
The system is governed by the incompressible Navier--Stokes equations, and we consider a Reynolds number $\Rey=2000$ based on the channel half-height and the centerline velocity, and a domain size of $2\pi\times 2\times 2\pi$ dimensionless length units along the $x$, $y$, and $z$ coordinates that indicate the streamwise, wall-normal, and spanwise directions, respectively. 
The state vector in this case is composed of the three-dimensional flow field of disturbances about the base parabolic velocity profile.

\begin{figure}
\centering
\includegraphics[width=1\textwidth]{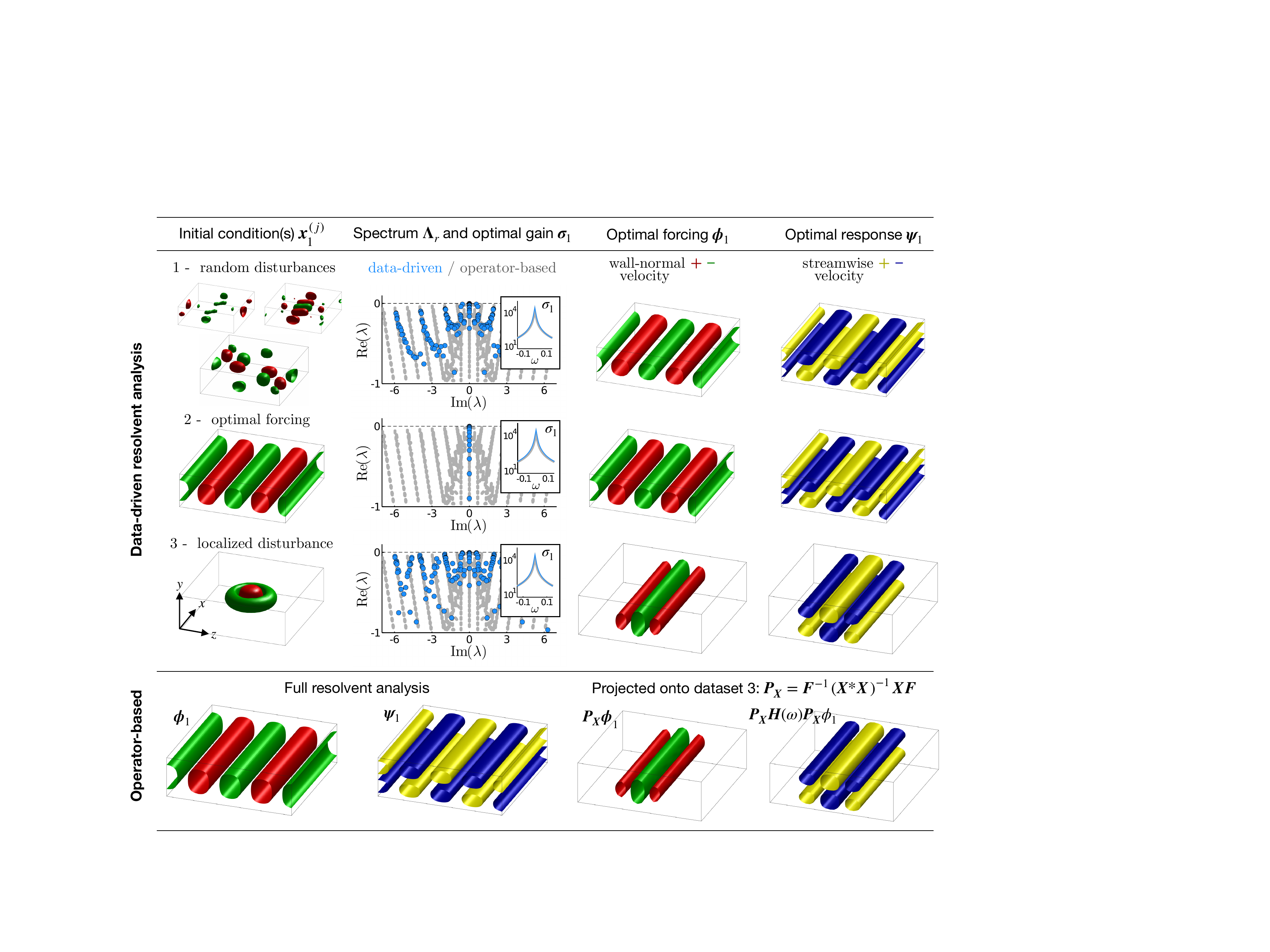}
\caption{Data-driven resolvent analysis of three-dimensional plane channel flow at $\Rey=2000$ based on the channel half-height and the centerline velocity. 
The method is demonstrated using three datasets obtained from DNS initialized with: $1$) small-wavenumber random disturbances, where three trajectories are considered, $2$) the optimal forcing, and $3$) localized actuation. Operator-based results are also shown for comparison, including the resolvent modes obtained when the input and output are constrained to lie in the span of the snapshots from dataset 3.}
\label{chflow}
\end{figure}

In the case of single-wavenumber perturbations, the dynamics are described by the traditional Orr-Sommerfeld and Squire equations \citep{Schmidbook}. 
The corresponding linear operator is built using Chebyshev spectral collocation \citep{Trefethen2000} to discretize the wall-normal direction. 
The Orr-Sommerfeld/Squire operator is then used to compute the operator-based spectrum and the leading resolvent gain of the three-dimensional flow, as shown in Fig. \ref{method}. 
This is achieved looping over wavenumber combinations that are compatible with the finite channel dimensions, i.e., integers for the current setup. We consider $N_y=101$ collocation points, and wavenumbers in the range $|\alpha| \le 7,$ for the streamwise component, and $|\beta| \le 7$ for the spanwise component. 
The operator-based leading resolvent modes, shown in Fig.~\ref{chflow}, are computed using $\omega	=0, \ \alpha=0, \ \beta=2$, for which the maximum gain is observed to occur. 
The optimal forcing and response correspond to the familiar streamwise vortices that excite streamwise streaks \citep{Trefethen1993}.

In order to demonstrate our data-driven resolvent analysis, we need snapshots from the transient evolution of the full three-dimensional flow field. 
We use the spectral code \emph{Channelflow} \citep{Gibson2008,Gibson2014} to perform direct numerical simulations of the incompressible Navier-Stokes equations. 
The code uses Chebyshev and Fourier expansions of the flow field in the wall-normal and horizontal directions, and a $3^{rd}$-order Adams--Bashforth backward differentiation scheme for the time integration. 
We find that a grid with $N_y=65$ and $N_x=N_z=32$ points is sufficient to discretize the domain for the cases studied, and a time step of $0.01$ time units is selected, keeping the CFL number at $0.32$. 
All cases described below are simulated for $500$ time units, and snapshots are saved every $\Delta t=0.5$ time units. 
The perturbation kinetic energy of all initial conditions simulated is set to $10^{-5}$ to ensure that the effect of nonlinearity is negligible.

In addition to \emph{how much} data is required for the data-driven resolvent, which was discussed in the previous section, here we investigate the more interesting question of \emph{which data}. 
To do so, we learn the leading resolvent modes from three datasets obtained from qualitatively different initial perturbations. 
First, we consider an initial perturbation of the wall-normal velocity component with the following distribution
\begin{equation}
v(x,y,z,0)=\sum_{\alpha,\beta}c_{\alpha} c_{\beta}\left(\cos(\pi y) +1\right)e^{\ii(\alpha x + \beta z)} + \text{c.c.}, \quad \alpha,\beta\in \lbrace -3, \ ..., \ 3\rbrace,
\end{equation}
where the $y$-dependence is selected to satisfy exactly the boundary conditions at the walls, $c_{\alpha}$ and $c_{\beta}$ are randomly sampled real numbers between $-1$ and $1$, and the horizontal velocity components are computed from the continuity equation. 
The first dataset is then composed of the snapshots from three simulations, each initialized with these random small-wavenumber disturbances. 
Data-driven resolvent analysis is applied retaining $r=200$ DMD eigenvectors, and the learned gains and modes are shown in Figure~\ref{chflow}.

The second dataset considers a single simulation started using the optimal forcing mode as the initial condition. In this scenario, the optimal gain and modes can be learned with only $r=20$ DMD eigenvectors, as shown in Figure~\ref{chflow}. 
It is interesting to look at the DMD eigenvalues, which for this case form a small subset of those learned from the first dataset, as shown in Figure~\ref{chflow}. 
In the previous case we learned the eigenvalues with largest real part, now we discover ones from a very specific subset of the complex plane. 
This highlights that, although the resolvent modes can be accurately represented by very few eigenvectors, the amount of data required to learn those eigenvectors that form an efficient basis is highly-dependent on the dynamic trajectories sampled. 
Moreover, this opens up exciting research directions, emphasizing the importance of finding principled disturbance designs to effectively probe dynamical systems.

Lastly, the third dataset, considers a localized disturbance as initial condition, which was previously studied by \citet{Ilak2008}. The exact form of the wall-normal velocity component is
\begin{equation}
v(x,y,z,0)=\left(1-\frac{r^2}{c_r^2}\right)\left(\cos(\pi y) +1\right)e^{\left(-r^2/c_r^2-y^2/c_y^2 \right)},
\end{equation}
where $r^2=(x-\pi)^2+(z-\pi)^2$, and the parameters are set to $c_r=0.7$ and $c_y=0.6$. 
This type of disturbance is close to what could be generated in experiments using a spanwise and streamwise periodic array of axisymmetric jets injecting fluid perpendicular to the wall. 
Data-driven resolvent analysis is performed using $r=200$ DMD eigenvectors, and the resulting optimal forcing and response modes do not resemble the operator-based ones, as shown in Figure~\ref{chflow}. 
In this scenario, the true resolvent modes of the system do not lie on the span of the learned DMD eigenvectors. 
More importantly, this occurs because some of the eigenvectors required to represent the true resolvent modes are not in the span of the data snapshots, and thus cannot be learned by DMD. 
In fact, we show in Figure~\ref{chflow} that the learned resolvent modes for this dataset coincide with the operator-based ones projected onto the span of the data snapshots, which is as good as we can expect to do with a data-driven approach. 

\section{Summary and conclusions}
\label{sec:conclusions}

In this work, we have developed an algorithm to perform resolvent analysis based on time-resolved data of dynamical systems. 
Unlike other modal decompositions, resolvent modes provide insight into the most amplified states, the most sensitive actuator locations, and the most responsive control inputs. 
Our method relies on dynamic mode decomposition to learn approximate eigenvalues and eigenvectors of the underlying linear operator from snapshots of transient dynamics of the system. 
Subsequently, we are able to compute the resolvent operator projected onto the learned eigenbasis. 
We perform data-driven resolvent analysis on numerical data of the linearized complex Ginzburg--Landau equation and of disturbances in a three-dimensional plane channel flow, demonstrating agreement between the leading data-driven and operator-based resolvent modes and gain distribution. 
A critical requirement is the design of initial disturbances to generate transients that are dynamically rich. 
We show that, using disturbances from a localized actuator, our method recovers the optimal forcing and response of the underlying system projected onto the span of the measured snapshots. 
This stresses the need for strategies to effectively explore the state space of a dynamical system. 

The proposed algorithm performs resolvent analysis in an equation-free and adjoint-free manner, therefore opening the possibility of only using experimental measurements. 
Data-driven resolvent analysis will play a significant role in lowering the barrier of entry to resolvent research and applications. 
Our results are encouraging for linearly stable flows; however, more work is required before applying this technique to turbulent flows, where the linear and nonlinear contributions to the transient dynamics of the system must be disambiguated.

\section*{Acknowledgements} 
This work was supported by U.S. Army Research Office (ARO W911NF-17-1-0306) and by the PRIME programme of the German Academic Exchange Service (DAAD) with funds from the German Federal Ministry of Education and Research (BMBF) and by the Deutsche Forschungsgemeinschaft (DFG) project number SE 2504/3-1.  

\newpage
\bibliographystyle{my_abbrvnat}
 \begin{spacing}{.9}
 \small{
 \setlength{\bibsep}{5.2pt}
 \bibliography{library}
 }
 \end{spacing}
\end{document}